# Aperiodic space-inhomogeneous quantum walks: localization properties, energy spectra and enhancement of entanglement

A. R. C. Buarque and W.S. Dias

*Instituto de Física, Universidade Federal de Alagoas, 57072-900 Maceió, Alagoas, Brazil*

We study the localization properties, energy spectra and coin-position entanglement of the aperiodic discrete-time quantum walks. The aperiodicity is described by spatially dependent quantum coins distributed on the lattice, whose distribution is neither periodic (Bloch-like) nor random (Anderson-like). Within transport properties we identified delocalized/localized quantum walks mediated by a proper adjusting of aperiodic parameter. Both scenarios are studied by exploring typical quantities (inverse participation ratio and survival probability), as well as the energy spectra of an associated effective Hamiltonian. By using the energy spectra analysis, we show that the early stage the inhomogeneity leads to vanishing gap between two main bands, which justifies the delocalized character observed for $\nu < 0.5$. With increase of $\nu$ arises gaps and flat-bands on the energy spectra, which corroborates the suppresion of transport detected for $\nu > 0.5$. For $\nu$ high enough we observe an energy spectra which resembles that described by the 1d Anderson model. Within coin-position entanglement, we show many settings in which an enhancement in the ability to entangle is observed. This behavior brings new informations about the role played by aperiodicity on the coin-poisition entanglement for static inhomogeneous systems, reported before as almost always reducing the entanglement when comparing with the homogeneous case. We extend the analysis in order to show that systems with static inhomogeneity are able to exhibit asymptotic limit of entanglement.

## I. INTRODUCTION

With different transport properties to its classical analog, quantum walks have been proven to be versatile and highly controllable platform to describe quantum systems [1–18] and quantum algorithms [19–24]. Besides the presence in different branches of science such as ferromagnetic films [3], bacteria in biological systems [4], quantum dots [5] and photosynthetic systems [7], quantum walks have also been the central subject of a wide range of experimental studies, either using nuclear magnetic resonance [11, 12], trapped atoms [13, 14], linear optics [15, 16] or integrated photonics circuits [17, 18].

Categorized in two classes, the evolution of the walker is determined completely by an unitary evolution. In continuous-time quantum walks the dynamics is described by a Hamiltonian which defines a progress on continuous time and discrete space, without coin-like degrees of freedom [13, 19]. In discrete-time quantum walks the walker propagates in discrete steps determined by a dynamic internal degree of freedom, which plays the role of a quantum coin [1, 9, 11]. The generating Hamiltonian is not needed, although it can be related to the procedure of integrating a certain Hamiltonian over a finite time [9, 11].

Restricting to disordered systems, continuous-time quantum walks have been studied for years, presenting today a well-established theoretical framework. The main character is the localization effect, where studies about the role played by system dimensionality [25–27], correlations [28–32] and electron-electron interaction [33–36], for example, were shown both in theoretical and experimental scope. More recently proposed and with more degrees of freedom, disordered discrete quantum walks need a better understanding, making it the reason of many studies on the theme. In this regime, the disorder may be associated to quantum coin or the displacement operator [15, 18, 37–54].

In general lines, the disorder/inhomogeneity induces deviations from quadratic spreading of the wave packet, including the emergence of Anderson localization. For a dynamic disorder, the quantum coin is the same at all lattice sites but changes at each time step [37–44]. In this context, the analysis of the time evolution of a quantum walker in the presence of unitary noise in the Hadamard operator shows that the standard deviation of the spatial distribution acquires a diffusive behavior for long times, like the classical random walk [37]. For quantum coins arranged in aperiodic sequences, like Fibonacci [39] and Thue-Morse [43], the transport properties of a quantum walker in a one-dimensional chain reveals a superdiffusive wave packet spreading. For Fibonacci distributions, this behavior has been connected with the power-law decay of the time-correlation function of the chaotic trace map [41]. By using a time dependence for the coin operator, different types of asymptotic behaviors (like subdiffusive and localized) for the wave-function spreading were reported [44].

A localized behavior has also been described for systems with spatial inhomogeneity, either for different coins randomly distributed along the lattice sites but fixed during the time evolution[15, 18, 45, 46, 52], as well as for systems with position-dependent phase defects [47–49, 54]. Both descriptions have experimental studies by using optical setups, which demonstrates the relevance of such structures for the advance of quantum information science [15, 18, 49]. The absence of localization was described for different quasiperiodic distributions in the inhomogeneity of the coin operators [43]. By using the distribution of two coins on the lattice, Fibonacci and the Thue-Morse sequences induce a superdiffusive behavior for the walker whereas the Rudin-Shapiro sequence shows a subdiffusive one [43]. A fractal nature was reported for quasiperiodic distributions in which the period of coin operators is distributed analogous to the Aubry-André model. The localized behavior is described for the inverse period of coins is an irrational number [53].

It is observed that many studies focus on associating the inhomogeneity implementation to transport properties of the walker. However, the coin-position entanglement has relevant aspects in quantum computation protocols and quantum information [55]. Contrary to the homogeneous quantum walk, systems with dynamical inhomogeneity create maximally coin-position entangled states in the asymptotic limit, independent of the initial condition of the walker [56]. On the other hand, the static inhomogeneity is almost always worse than the homogeneous quantum walk in terms of entanglement generation [45]. In addition, it was reported an absence of asymptotic limit for such systems [45]. The decrease (enhancement) of entanglement due to static (dynamic) inhomogeneity has also described for 1d systems with on-site phase disorder [47]. The effect of delocalization of the initial state shows a relation between the initial angles of the spin amplitudes, which always leads to the maximal entanglement for Hadamard and Fourier walks [57].

Faced with the above descriptions, we study space-inhomogeneous discrete quantum walks which exhibit transitions between localized and delocalized regimes. Here, the inhomogeneity is described by a deterministic distribution of quantum coins along the lattice, whose characteristic is neither periodic (Bloch-like) nor random (Anderson-like). The model proposed here may be seen as a generalization of the model investigated in ref. [53], whose description for the coin operator can be restored by a proper adjust of the aperiodicity present here. Transport properties are studied by exploring typical quantities (inverse participation ratio and survival probability), as well as the energy spectra of an associated effective Hamiltonian. Localized and delocalized quantum walks are mapped, which seems useful to control quantum information and quantum processing. We also study the coin-position entanglement. The role played by correlations on the coin-position entanglement, as well as the ability of systems with static inhomogeneity to create maximally entangled states or have an asymptotic limit were issues explored in our study.

## II. MODEL

We consider a quantum walker moving in an infinite 1D lattice of interconnected sites. The walker consists of a qubit (two-state quantum system) with the internal degree of freedom (e.g. spin[1] or polarization [15]). The quantum walker state $|\psi\rangle$ belongs to a Hilbert space $H = H_c \otimes H_p$, where $H_c$ is a complex vector space of dimension 2 associated with the internal degree of freedom of the qubit, and $H_p$ denotes a countable infinite-dimensional space associated to site positions. Here, the internal degree of freedom of the walker (coin) is spanned by the orthonormal basis $\{|\uparrow\rangle, |\downarrow\rangle\}$, while the position space is spanned by the orthonormal basis $\{|n\rangle : n \in \mathbb{Z}\}$. Thus, a general initial state ($t = 0$) can be written as

$$|\psi(0)\rangle = \sum_n [a(n,0)|\uparrow\rangle \otimes |n\rangle + b(n,0)|\downarrow\rangle \otimes |n\rangle], \quad (1)$$

with $\sum_n |a(n,0)|^2 + |b(n,0)|^2 = 1$.

The evolution of system depends on both internal and spatial degree of freedom. We start the quantum walker in the initial position and act upon it with a unitary operator $\hat{C}$, well known as quantum coin or quantum gate, followed by a conditional displacement operation $\hat{S}$ at each time step. That is, the position of a particle evolves according to its internal coin state. Thus, the state of the walker after $t$ steps is given by applying the unitary transformation $|\psi(t)\rangle = \hat{U}^t |\psi(0)\rangle$ to the initial state, where $\hat{U}^t = \prod_{i=1}^{t} \hat{S} \cdot (\hat{C}_i \otimes \mathbb{I}_p)$. The displacement operator has the form

$$\hat{S} = \sum_n (|\uparrow\rangle\langle\uparrow| \otimes |n+1\rangle\langle n| + |\downarrow\rangle\langle\downarrow| \otimes |n-1\rangle\langle n|), (2)$$

while $\mathbb{I}_p$ is the identity operator defined over $H_p$.

Here, we introduce a spatial inhomogeneity in the coin operators given by an aperiodic distribution

$$\hat{C}_n = \cos(\theta_0 n^\nu)|\uparrow\rangle\langle\uparrow| + \sin(\theta_0 n^\nu)|\uparrow\rangle\langle\downarrow| \\ + \sin(\theta_0 n^\nu)|\downarrow\rangle\langle\uparrow| - \cos(\theta_0 n^\nu)|\downarrow\rangle\langle\downarrow|. \quad (3)$$

Thus, the coin operators depend on their positions $n$ and $\nu$ is a tunable parameter that controls the aperiodicity degree on the rotation angle $\theta_0 = [0, 2\pi]$. The model proposed here may be seen as a generalization of the model investigated in ref. [53], where its description for the coin operator can be restored by setting $\nu = 1$. On the other hand, by setting $\nu = 0$, we recover the homogeneous quantum walk. Thus, the $\nu \neq 0$ parameter induces a static inhomogeneity on the system with position dependent quantum coins operations fixed during the time evolution, i.e., $\hat{C}(\theta_0, n, t) = \hat{C}(\theta_0, n) = \hat{C}_n(\theta_0)$.

## III. RESULTS AND DISUSSION

### A. Transport properties and energy spectra

Initially, we investigate the influence of aperiodicity on the transport features of walker. Thus, we start following the time evolution of the probability density distribution for a walker whose initial state is $|\psi(0)\rangle = |\uparrow\rangle \otimes |n_0\rangle$, with $n_0 = 40$. In order to better understand the influence of aperiodic spatial inhomogeneity on the quantum walk, we choose two well-known quantum coins as reference: Hadamard - $\theta_0 = \pi/4$; and Pauli-X - $\theta_0 = \pi/2$. While Hadamard coins uniformly distributed on the lattice sites induce a spread of the probability distribution (fig. 1a), a spatial inhomogeneity with $\nu = 1.0$ gives rise to localized behavior whose probability distribution remains restricted around the initial sites (fig. 1b). On the other hand, whereas Pauli-X coins brings a localized behavior whose probability distribution stays oscillating around the initial site (fig. 1c), a spatial inhomogeneity with $\nu = 0.05$ leads to a delocalized distribution (fig. 1d).

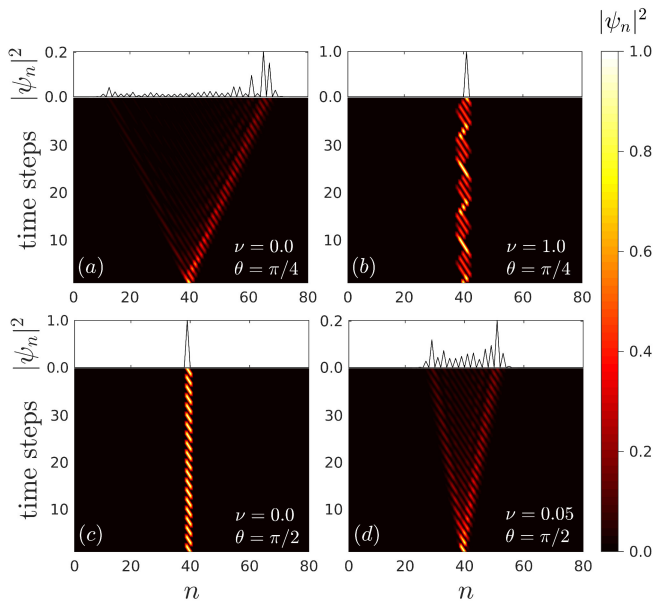

FIG. 1. (Color on-line) Time evolution of the density of probability in position space of a quantum walker with initial state $|\psi(0)\rangle = |\uparrow\rangle \otimes |40\rangle$ and ruled by (a) $\theta_0 = \pi/4$ and $\nu = 0.00$; (b) $\theta_0 = \pi/4$ and $\nu = 1.00$; (c) $\theta_0 = \pi/2$ and $\nu = 0.00$; (d) $\theta_0 = \pi/2$ and $\nu = 0.05$. Specific spatial inhomogeneities on well-known quantum coins (Hadamard and Pauli-X) induce opposite dynamics those shown by homogeneous systems.

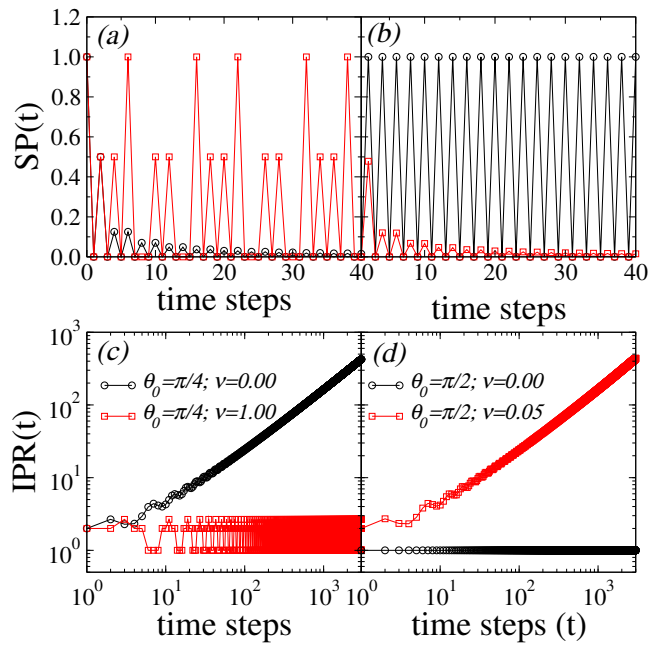

FIG. 2. (Color on-line) Time evolution of (top panels) survival probability (SP) and (bottom panels) the inverse participation ratio (IPR) for same configurations of $\theta_0$ and $\nu$ used in fig. 1: (a-c) $\theta_0 = \pi/4$ with $\nu = 0.00$ and $\nu = 1.00$; (b-d) $\theta_0 = \pi/2$ with $\nu = 0.00$ and $\nu = 0.05$. Both quantities ratify the transitions between delocalized and strong-localized regimes ruled by aperiodic inhomogeneity.

The above dynamics can be better characterized by computing the survival probability

$$\text{SP}(t) = \sum_{\sigma=\uparrow,\downarrow} |\langle n| \otimes \langle \sigma | \psi(t)\rangle|^2 \Big|_{n=n_0} \qquad (4)$$

and the inverse participation ratio

$$\text{IPR}(t) = \frac{1}{\sum_n |\psi_n(t)|^4}. \qquad (5)$$

The first describes the probability that the walker returns to the initial position at time $t$, whereas IPR(t) gives the estimate number of sites over which the wave packet is spread at time $t$. Thus, in the long-time regime the survival probability saturates at a finite value for a localized wave function, while $\text{SP}(t) \to 0$ means that the walker escapes from its initial location. On the other hand, in the long-time regime the $\text{IPR}(t) \propto N^0$ indicates that the walker remains localized, whereas $\text{IPR}(t) \propto N$ corresponds to the regime where the wave function is distributed over the lattice. With this analysis in mind, we computed both quantities for the same configurations used in fig.1 and we observe the results reinforcing the previous description (see fig.2). For systems ruled by Hadmard coins homogeneously distributed on the lattice the $\text{SP}(t) \to 0$ and IPR(t) grows linearly as time evolves. A localized behavior, where the wave function is split into equiprobables portions and grouped into its initial position, describes the dynamics for an inhomogeneous case of $\theta_0 = \pi/4$ and $\nu = 1.00$. For homogenous systems of Pauli-X coins a bit-flip dynamics is well characterized, with SP(t) successively alternating between 0 and 1 and IPR(t) remaining fixed in 1. When we tune $\nu = 0.05$ for $\theta_0 = \pi/2$ we observe $\text{SP}(t) \to 0$ and IPR(t) growing linearly as time evolves.

To understand the origin of changes in the mobility of the walker we introduce an associated effective Hamiltonian and explored the nature of its eigenvalues. As described above, the dynamics of the quantum walker is given by a unitary time-step operator, which consists in a change of internal degree of freedom ("coin flip") followed by a coin-dependent displacement operation:

$$|\psi(t+1)\rangle = \hat{U}|\psi(t)\rangle. \qquad (6)$$

Since $\hat{U}$ is time-independent, by choosing the time unit as the period of the time evolution and the unit of position as the period of the lattice, $\hat{U}$ can be interpreted as a Floquet operator [46, 51] associate with a time-independent

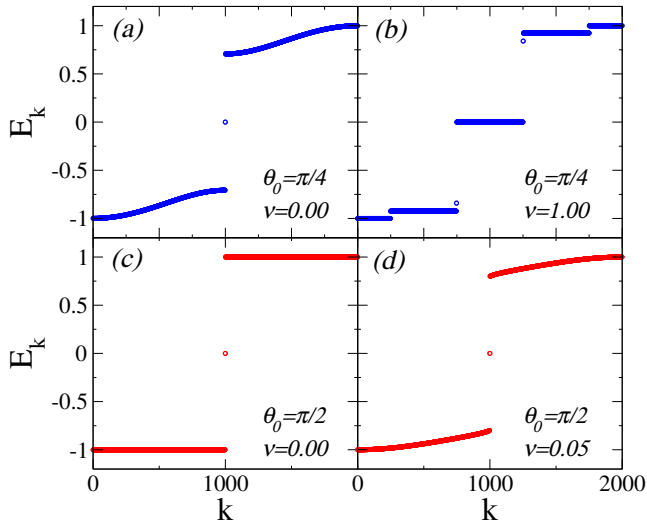

FIG. 3. Energy spectrum $E_k = -ilog(\lambda_k)$, with $\lambda_k$ being the eigenvalues of the operator $\hat{U}$ for same configurations used in fig. 1. Allied to the presence of the internal degree of freedom, both systems exhibit two main bands in absence of inhomogeneity: Hadamard coins shows a continuous energy spectrum inside the two main bands, while Pauli-X coins shows a flat degenerate energy spectrum into both bands. The inhomogeneity for $\theta_0 = \pi/4$ open new gaps in the main energy bands, besides imposing a degeneracy on them. On the other hand, the inhomogeneity breaks the high degeneracy level within two sub bands for $\theta_0 = \pi/2$.

effective Hamiltonian ($H_{eff}$) defined as

$$\hat{U} = e^{-iH_{eff}}. \tag{7}$$

Thus, by using the same methodology employed in ref. [43], we compute $E_k = -i\log(\lambda_k)$, where $\lambda_k$ are the eigenvalues of the operator $\hat{U}$. In figure 3 we explore the energy spectra of same configurations shown in previous figures. In general lines, homogeneous systems exhibit two main bands, which is related to the presence of the internal degree of freedom (coin). For homogeneous systems composed by Hadamard coins ($\theta_0 = \pi/4$) we observe two main bands, gapped away from each other, but with continuous energy spectrum inside them. On the other hand, Pauli-X coins ($\theta_0 = \pi/2$) uniformly distributed gives a flat degenerate energy spectrum for both bands, so that the linear combination of eigenstates results in the well known bit flip dynamics [55] exhibited in fig. 1c and fig. 2b,d. Both descriptions are in agreement with previous literature [43, 46]. In fig. 3b we observe that the spatial inhomogeneity of $\nu = 1.0$ for $\theta_0 = \pi/4$ induces the emergence of new gaps within the main energy bands, giving rise new sub bands. The eigenstates are then still compactly localized inside these sub bands, so as to exhibit a degenerate aspect. Thus, a linear combination of eigenstates results in a localized dynamics restricted to a few sites. On the other hand, the spatial inhomogeneity of $\nu = 0.05$ for $\theta_0 = \pi/2$ breaks the high degeneracy within main bands. Thus, the inhomogeneity of $\theta_0$ induces an enlargement of bands and decreases of gap between them. These aspects favors the spread of wave function along the lattice, as seen in fig.1d and fig. 2b,d. The spatial inhomogeneity altering the dispersion and gaps of energy corroborates the association of $\theta_0$ to the kinect-energy made in ref. [46].

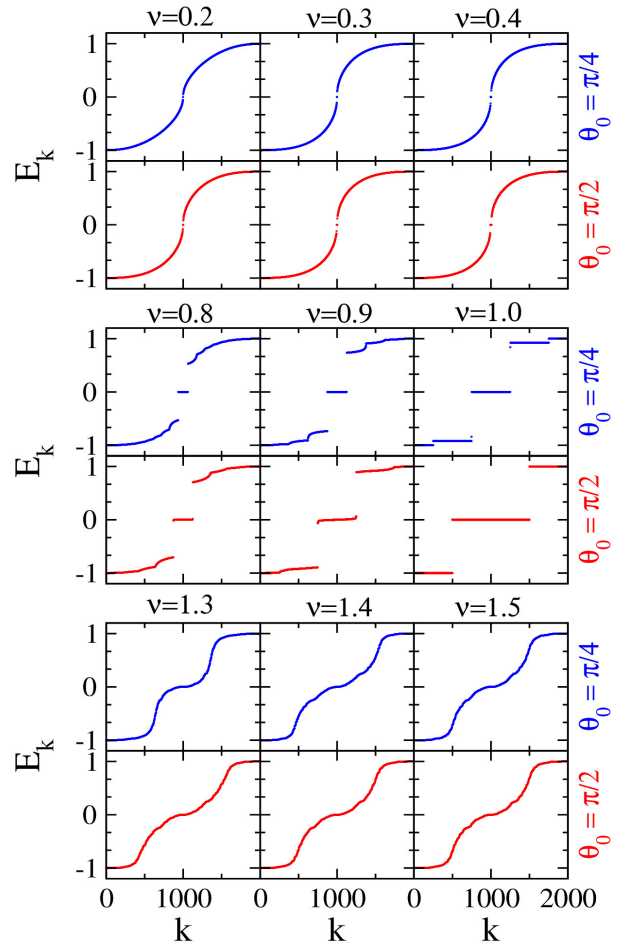

FIG. 4. (Color online) Energy spectrum $E_k = -ilog(\lambda_k)$, with $\lambda_k$ being the eigenvalues of the operator $\hat{U}$ for $\theta_0 = \pi/4$ (blue) and $\theta_0 = \pi/2$ (red). Top panels: $\nu = 0.2, 0.3, 0.4$; middle panels: $\nu = 0.8, 0.9, 1.0$; bottom panels: $\nu = 1.2, 1.3, 1.4$. In the early stage the inhomogeneity leads to vanishing gap between two main bands. With increase of $\nu$ it is observed the presence of gaps and flat-bands structures. The adjustment of $\nu > 1$ provides a irregular spectrum which resembles that described by 1d Anderson model.

A more extensive description about the relation between the energy spectrum and dynamics behavior of inhomogeneous quantum walks is shown fig. 4, where we compute $E_k$ for both coins ($\theta_0 = \pi/2, \pi/4$) with setting of $\nu = 0.2, 0.3, 0.4$, $\nu = 0.8, 0.9, 1.0$ and $\nu = 1.2, 1.3, 1.4$. In the early stage the inhomogeneity induces the width of each band to its maximal value, which leads to vanishing gap, i.e. a gapless spectrum. This nature remember the spectrum obtained from homogeneous Pauli Z coins ($\theta_0 = 0$), whose delocalized character is well-known [55]. With increase of $\nu$ arises new gaps, subbands and degenerate eigenstates. The emergence of flat-band states can be understood as states located within a small part of the lattice. With the energy spectrum independent of momentum, the kinetic energy is quenched, the group velocity vanishes and the suppresion of transport is observed. For $\nu$ high enough the energy spectra which resembles that described by the 1d Anderson model are observed. This analysis is consistent with to arrangement of the quantum coins along the lattice, whose slowly varying distribution exhibited for small $\nu$ becomes random-like as $\nu$ increases.

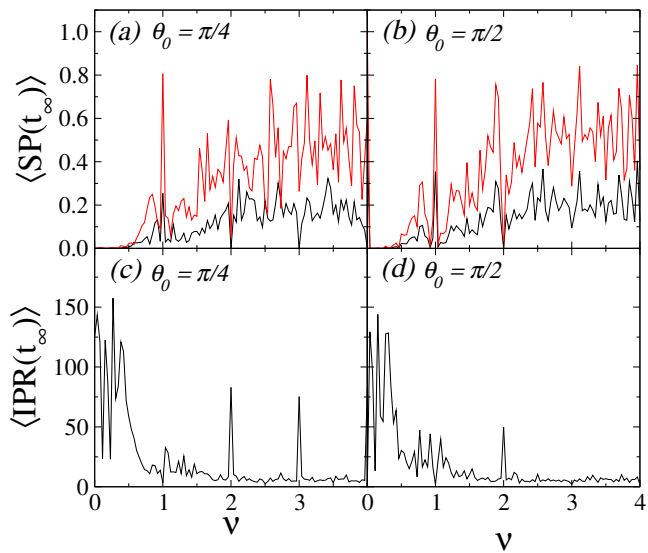

FIG. 5. (Color on-line) Long-time average of the survival probability ($\langle SP(t_\infty) \rangle$) and the inversal participation ratio ($\langle IPR(t_\infty) \rangle$) versus $\nu$, for $\theta_0 = \pi/2, \pi/4$. We display for $\langle SP(t_\infty) \rangle$ the average probability of the walker returns to the initial (bottom/black) and around the initial position (top/red) - see descripton in text. Delocalized (localized) quantum walks are predominant for $\nu < 0.5$ ($\nu > 0.5$), consistent with energy spectra analysis. The translational symmetry on the inhomogeneity ($\nu \in \mathbb{Z}^*$) favors the spread of wave function along the lattice.

The above analysis suggests the $\nu$ parameter as able to tune delocalized and localized quantum walks, such that the delocalized character is predominant in the low-inhomogeneity regime. In order to confirm, we explore in fig. 5 the long time average of survival probability ($\langle SP(t_\infty) \rangle$) and the inversal participation ratio ($\langle IPR(t_\infty) \rangle$) for both coins ($\theta_0 = \pi/2, \pi/4$), by ranging $\nu$ from 0 to 4. As before, we follow considering an asymmetric initial state $|\psi(0)\rangle = |\uparrow\rangle \otimes |n_0\rangle$, with $n_0 = 500$. Besides the average probability of the walker returns to the initial state ($n = n_0$ in eq. 4) after a long time (bottom/balck), we show in fig.5a,b the average probability of finding the walker around ($n = n_0, n_0 \pm 1$ in eq. 4) the initial position (top/red). Delocalized quantum walks are well-described for both coins in low-inhomogeneity regime ($\nu < 0.5$). Besides $\langle SP(t_\infty) \rangle \to 0$, we observe $\langle IPR(t_\infty) \rangle$ predominantly high. Some configurations with low $\langle IPR(t_\infty) \rangle$ are related to delocalized quantum walks whose probability density is restricted to few sites, similar to the walker subjected to homogeneously distributed Pauli-Z coins [38, 55]. The localized nature is predominant for higher $\nu$ values. By using a connection between discrete and continuous quantum walks [58], we interpret this aperiodic inhomogeneity (within a low $\nu$ regime) as a potential difference between neighboring sites proportional to $n^{\nu-1}$, which vanishes in the thermodynamic limit and favors a broading of walker. For abrupt variations on the distribution of quantum coins the internal correlations becomes effectively short-ranged, promoting destructive interferences and inducing a localized quantum walk. This premise is reinforced by delocalization of wave function around $\nu \in \mathbb{Z}^*$, when inhomogeneity on the quantum coins distribution recovers a translational symmetry. However, the spatial dependence imposes intrinsic correlations on the interference terms of the quantum coins, making the translational symmetry not always predominant over the delocalization criterion.

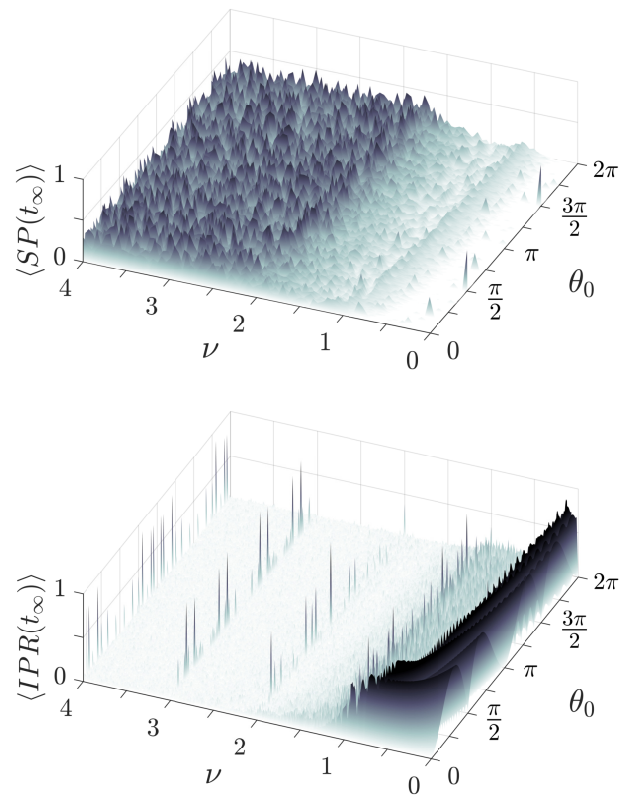

FIG. 6. (Color on-line) Diagram of long-time average of the survival probability ($\langle SP(t_\infty) \rangle$) and the inversal participation ratio ($\langle IPR(t_\infty) \rangle$) versus $\nu$ versus $\theta_0$. Delocalized quantum walks are predominant for $\nu < 0.5$, while the localized nature is predominant with increasing $\nu$, except for $\nu \in \mathbb{Z}^*$. The latter, which may be associated with the translational symmetry of inhomogeneity, appear to be a relevant aspect about the spread of wave function along the lattice.

An more extensive description about the dynamics behavior of quantum walker is shown in fig. 6, where we display diagrams for long-time average of the survival probability ($\langle SP(t_\infty) \rangle$) and the inversal participation ratio ($\langle IPR(t_\infty) \rangle$) versus $\nu$ versus $\theta_0$. Here, we consider the maximal IPR between collected data in order to plot on the vertical axis a normalized $\langle IPR(t_\infty) \rangle$. The periodic dependence exhibited for $\nu = 0$ (disorder-free) describes the well known behavior for different quantum coins: For Pauli-Z coins the quantum walker will move away from initial position and the wave function will only be seen at the position $\pm t$ with non-zero probability. On the other hand, for Pauli-X coins the quantum walker will remain localized around initial position for all time t. By including a spatial dependence ($\nu \neq 0$), the previous description is changed. As previously suggested, delocalized (localized) quantum walks are predominant for $\nu < 0.5$ ($\nu > 0.5$). The translational symmetry on the inhomogeneity, recovered by $\nu \in \mathbb{Z}^*$, appear to be a relevant aspect about the spread of wave function along the lattice.

### B. Entanglement properties

By considering the changes on the spreading behavior of the quantum walker described previously, we investigate the entanglement between the coin state and the particle position. In this scenario, where the evolution of the whole system is given by an unitary transformation in which the coin operation at each step directly controls the interference pattern of the quantum walker, we comput the von Neumann entropy of the reduced density matrix given by [45, 47, 55–57]

$$S_E(t) = -Tr\left[\rho_c(t) \log_2 \rho_c(t)\right] \quad (8)$$

where $\rho_c$ is the reduced density matrix obtained by tracing over the position degree of freedom the full density matrix $\rho = |\psi(t)\rangle\langle\psi(t)|$ of the quantum walk system. Within the most general quantum state at some arbitrary step $t$

$$|\psi(t)\rangle = \sum_n \left[a_n(t) |\uparrow\rangle + b_n(t) |\downarrow\rangle\right] \otimes |n\rangle, \quad (9)$$

the reduced density matrix is given by

$$\rho_c(t) = \sum \langle m|\rho|m\rangle = \sum_m \begin{bmatrix} \alpha(t) & \gamma(t) \\ \gamma^*(t) & \beta(t) \end{bmatrix}, \quad (10)$$

with $\alpha(t) = \sum_m |a_m(t)|^2$, $\beta(t) = \sum_m |b_m(t)|^2$, $\gamma(t) = a_m(t)b_m^*(t)$ and the probability distribution $|\psi_n(t)|^2 = \alpha(t) + \beta(t) = 1$. By diagonalizing $\rho_c(t)$ we obtain

$$S_E\left[\rho_c(t)\right] = -\lambda_+ \log_2 \lambda_+ - \lambda_- \log_2 \lambda_-, \quad (11)$$

where $\lambda_\pm$ are the eigenvalues of matrix $\rho_c$,

$$\lambda_\pm = \frac{1}{2}\left\{1 \pm \sqrt{1 - 4\left[\alpha(t)\beta(t) - |\gamma(t)|^2\right]}\right\}. \quad (12)$$

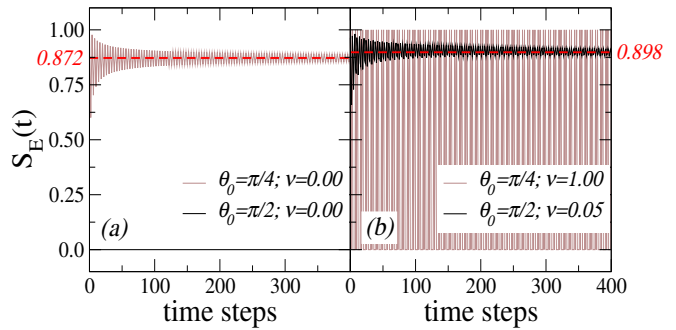

FIG. 7. (Color on-line) Time evolution of von Neumann entropy for same configurations of fig. 1, i.e. Hadamard and Pauli-X coins in (a) absence and (b) presence of spatial inhomogeneity. Whitin a disorder-free ($\nu = 0.0$) system, Hadamard coins induce a coin-position entanglement which saturates around 0.872, in fully agreement with ref. [59, 60]. On the other hand, Pauli-X coins shows absence of entanglement. Aperiodic inhomegeneity can alter significantly the entanglement properties.

Thus, $S_E(t) \in [0, 1]$, with separable states (not entangled) giving $S_E = 0$ and maximally entangled states providing $S_E = 1$. For a disorder-free system ruled by Hadamard coins with local initial conditions it has been shown whitin an asymptotic regime that $S_E(t \to \infty) \approx 0.872$ [59, 60].

Taking into account the above results, we start by computing the time evolution of von Neumann entropy $S_E(t)$ for a walker with initial state $|\psi(0)\rangle = |\uparrow\rangle \otimes |n_0\rangle$, $n_0 = 500$, subjected to the same configurations shown in fig. 1, i.e. $\theta_0 = \pi/4$ with (a) $\nu = 0.0$ and (b) $\nu = 1.0$, as well as $\theta_0 = \pi/2$ with (a) $\nu = 0.0$ and (b) $\nu = 0.05$. The results in fig. 7 reflect the localized/delocalized transition observed in fig. 1. For homogeneous case, Hadamard coins induce a coin-position entanglement which saturates around 0.872, in fully agreement with ref. [59, 60]. By adding spatial inhomogeneity ($\nu = 1.0$) the coin-position entanglement exhibits an oscillatory behavior with values $0 \leq S_E(t) \leq 1$, i.e. between no entangled and full entangled. This feature is consistent with dynamics behavior described in fig. 1 and the energy spectrum composed by flat-bands exhibited in fig. 3. On the other hand, while homogeneous $\theta_0 = \pi/2$ coins shows absence of entanglement for homogeneous distribution, an aperiodicity with $\nu = 0.05$ brings an entanglement which oscillates close 0.898. This latter feature comes from the emergence of dispersion on the energy spectrum, as described in fig. 3.

It has been reported that static inhomogeneity exhibits the worst results in terms of entanglement generation, almost always reducing the entanglement when comparing with the homogeneous case [45, 47]. In face of significant changes on the entanglement power induced by aperiodic spatial dependence (previously reported in fig. 7), we investigate whether this type of static inhomogeneity is able to make it more prone to increase the coin-position entanglement. In fig. 8 we focus on settings which exhibits a delocalized character (see top panels). We consider $\theta_0 = \pi/6, \pi/4, \pi/3$, where systems with spatial inhomogeneity ($\nu = 0.05$) are shown on the right panels. We observe an increasing of entanglement power induced by inhomegeneity, even for a configuration that has a strong entanglement into a homogeneous regime ($\theta_0 = \pi/3$). Such ability has no been reported for static inhomogeneity, only for dynamic and fluctuating inhomogeneities [45, 47]. Since both studies describe random distributions, for the quantum coins [45] or on-site phase [47], we identified that inner correlations on the distribution of disorder can enhance the efficiency in the ability to entangle internal (spin/polarization) and external (position) degrees of freedom.

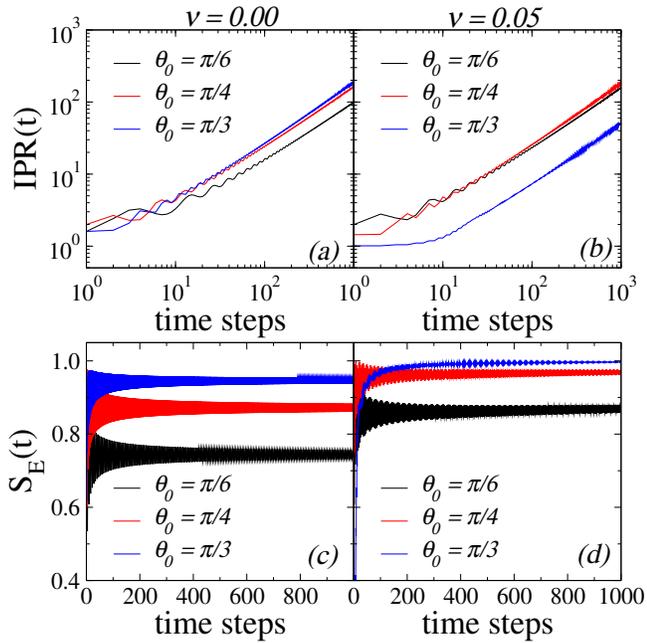

FIG. 8. (Color on-line) Time evolution of the inverse participation ration IPR(t) (top panels) and von Neumann entropy $S_E(t)$ (bottom panels) for coins $\theta_0 = \pi/6, \pi/4, \pi/3$ in the (a,c) absence and (b,d) presence of inhomogeneity. While IPR(t) exhibits a delocalized character of coins in both regimes, we observe an increasing of entanglement power induced by inhomegeneity, even for a configuration with a strong entanglement in the homogeneous configuration ($\theta_0 = \pi/3$).

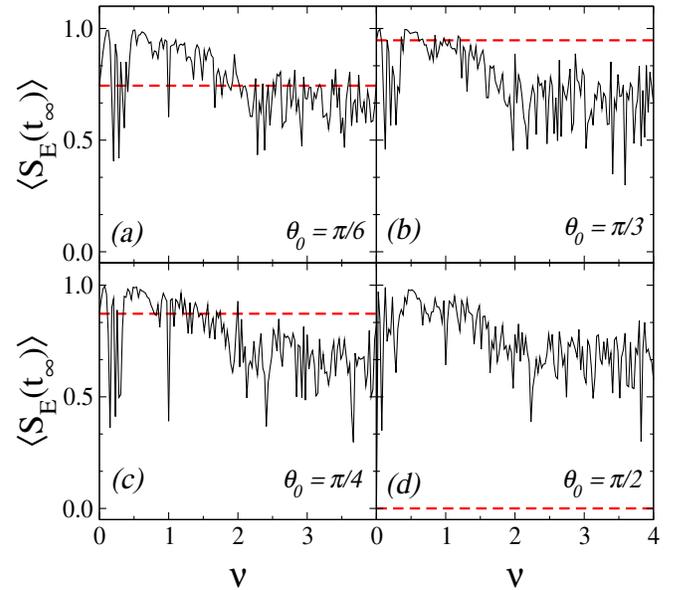

FIG. 9. The long-time average of the entanglement ($\langle S_E(t_\infty)\rangle$) versus $\nu$, for $\theta_0 = \pi/6, \pi/4, \pi/3$ and $\pi/2$. Dashed lines describe the respective $\langle S_E(t_\infty)\rangle$ in homogeneous systems. We observe many configurations of inhomogeneity in which there is an enhancement in the entanglement power, being predominantly found for the first values of $\nu$. As $\nu$ grows, the random-like aspect that governs the quantum coins distribution impels the entanglement to weakening.

In order to better understand this phenomenology we compute in the fig. 9 the long-time average of the entanglement ($\langle S_E(t_\infty)\rangle$) versus $\nu$, for $\theta_0 = \pi/6, \pi/4, \pi/3$ and $\pi/2$. To establish reference we show dashed lines that describe the $\langle S_E(t_\infty)\rangle$ for the respective homogeneous system. We observe many configurations of inhomogeny in which there is an enhancement on the entanglement power. This configurations are predominantly found for the first values of $\nu$. As $\nu$ grows, the random-like aspect that governs the quantum coins distribution impels the coin-position entanglement to weakening, which agrees with description of ref [45, 47].

By considering the raise in the entanglement power an important aspect, we extend our numerical experiments in order to offer a diagram $\theta_0$ versus $\nu$ which reveals this enhancement in the ability to entangle internal (spin/polarization) and external (position) degrees of freedom (see fig. 10). Here, we do not regard about the increase percentage, only with the grow of the long-time average entanglement as compared with its respective value for homogeneous case. Thus, black points mean any increase in long-time average of entanglement, while white points denote a weakening. For $\theta_0 = \pi/2$ and $\theta_0 = 3\pi/2$ we observe a continuous black line, since homogeneous systems of Pauli-X coins display $\langle S_E(t_\infty)\rangle = 0$. This latter is related to energy spectrum composed by two main flat-bands which induces a bit-flip dynamics. The increase in entanglement concentrates around Pauli-Z coins, configurations in which the walker moves away of initial position without interference and also shows absence of entanglement within a homogeneous regime. For initial $\nu$ values the gain of entanglement power is predominant, while becomes restricted to close values of $\theta_0 = 0, \pi$ and $2\pi$ as $\nu$ grows.

## C. Asymptotic limit

A complementary analysis was directed to asymptotic limit of entanglement. It has been reported that systems ruled by Hadamard quantum coins [60], systems with dynamic [45, 56] and fluctuating [45] inhomogeneities display an asymptotic limit. On the other hand, systems with static inhomogeneity has no asymptotic limit [45], with the long-time entanglement fluctuating about a mean value with no signs of convergence to its mean. Thus, by using the same procedure employed in ref. [45], we compute the trace distance

$$D\left[\rho_c(t+1), \rho_c(t)\right] = D(t) = \frac{1}{2}Tr\left|\rho_c(t) - \rho_c(t-1)\right|. \tag{13}$$

in order to study the asymptotic limit of our aperiodic quantum walks. This quantity gives how close are two quantum states, which in the dynamic context can be played as how well information is preserved by some physical process. Thus, whenever a quantum walk leads to

$$\lim_{t \to \infty} D\left[\rho_c(t+1), \rho_c(t)\right] = D(t) = 0, \tag{14}$$

it is said that the system has an asymptotic limit.

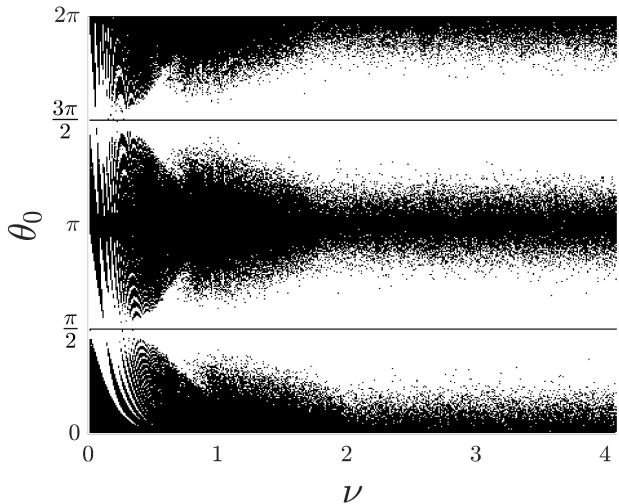

FIG. 10. Diagram $\theta_0$ versus $\nu$ which reveals the enhancement in the ability to entangle internal (spin/polarization) and external (position) degrees of freedom with respect to homogeneous distribution of quantum coins. We do not regard about an gain percentage, so that black points means an increase of long-time average of entanglement, while white points denote a weakening. The increase in entanglement is predominant for initial values of $\nu$. As $\nu$ grows, the increasing is restricted to values close to $\theta_0 = 0, \pi$ and $2\pi$, as well as $\theta_0 = \pi/2$ and $3\pi/2$.

In fig. 11 we show the results of this analysis for Hadamard coins ($\theta_0 = \pi/4$), Pauli-X ($\theta_0 = \pi/2$) and $\theta_0 = \pi/3$. Homogeneous systems of Hadamard coins show an asymptotic limit which obeys a power law dependence in agreement with ref. [45]. However, we found $D(t) \sim t^{-1.0}$, in contrast with $D(t) \sim t^{-1/2}$ reported by ref. [45]. We attribute this difference to the initial conditions, which has been considered here as localized initial condition ($|\psi(0)\rangle = |\uparrow\rangle \otimes |0\rangle$), while it was built through an average of 16.384 delocalized initial conditions (superposition between positions $|-1\rangle$ and $|1\rangle$) in ref. [45]. We also note $\theta_0 = \pi/4$ tuned with spatial dependence $\nu = 1.0$ performing quantum walks that does not come close to a particular state and do not have an asymptotic limit. For Pauli-X coins the asymptotic limit display opposite trends: while it has no asymptotic limit in homogeneous case, with an aperiodic inhomogeneity ruled by $\nu = 0.05$ it exhibits an asymptotic limit which obeys a power law dependence $D(t) \propto t^{-1.0}$. Furthermore, we can see the inhomogeneous case of Pauli-X coins displaying a linear coefficient slightly lower than that displayed by homogeneous Hadamard coins. Thus, we report a possibility to induce an asymptotic limit even in static inhomogeneity.

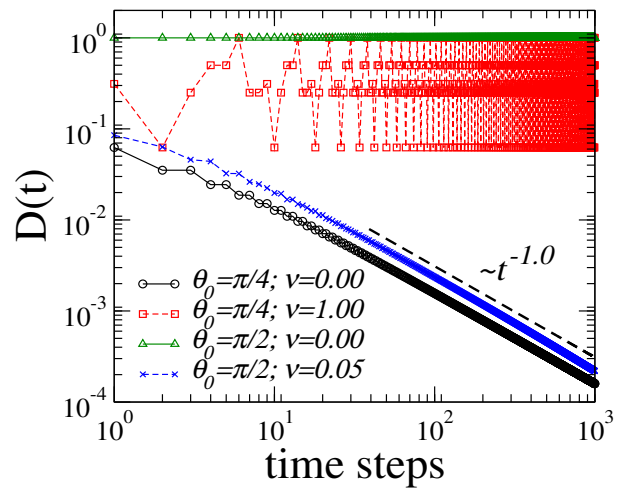

FIG. 11. (Color on-line) Time evolution of trace distance for $\theta_0 = \pi/4, \nu = 0.00$, $\theta_0 = \pi/4, \nu = 1.00$, $\theta_0 = \pi/2, \nu = 0.00$ and $\theta_0 = \pi/2, \nu = 0.05$. Like in systems governed by Hadamard coins, systems with static inhomogeneity can also exhibit asymptotic value which obeys a power law. Dashed line is a guide to the eyes showing $D(t) \sim t^{-1.0}$

## IV. SUMMARY AND CONCLUSIONS

In summary, we have studied the localization and entanglement properties in quantum walks ruled by aperiodic (inhomogeneous) spatial dependence on the quantum coins distribution. With an adjustable distribution ruled by a single parameter $\nu$ we show the existence of delocalized and localized quantum walks, as well as the proper adjusting of the aperiodicity in order to developing both. With the energy spectra analysis, obtained from an associated effective Hamiltonian and the nature of its eigenvalues, both delocalized and localized regimes could be better understood. In the early stage the inhomogeneity leads to vanishing gap between two main bands, which justifies the delocalized behavior observed for $\nu < 0.5$. With increase of $\nu$ arises gaps and flat-bands on the energy spectra, which justifies the suppresion of transport detected for $\nu > 0.5$. For $\nu$ high enough the energy spectra resembles that described by the 1d Anderson model. The translational symmetry on the inhomogeneity, recovered for $\nu \in \mathbb{Z}^*$, shows to be a relevant aspect, which favors delocalized quantum walks. For the coin-position entanglement, taking as reference the homogeneous distribution (disorder-free) of quantum coins, we identified many settings in which an enhancement in the ability to entangle is observed. This behavior brings new informations about the role played by aperiodicity on the coin-poisition entanglement for static inhomogeneous systems, reported before as almost always reducing the entanglement when comparing with the homogeneous case [45, 47]. Furthermore, since systems with static inhomogeneity has been reported as having no asymptotic limit [45], we extend the analysis in order to show that aperiodic spatial inhomogeneity is able to induces an asymptotic limit to entanglement. To conclude, with the recent experimental achievements in optical setups [18], we believe that the scheme proposed here is feasible for integrated waveguide circuits. The spatial inhomogeneity proposed here would be ajusted in the beamsplitters arranged in a lattice of Mach-Zehnder interferometers, since each beamsplitter implements the quantum coin operation.

## V. ACKNOWLEDGMENTS

This work was partially supported by CNPq (The Brazilian National Council for Scientific and Technological Development), CAPES (Federal Brazilian Agency) and FAPEAL (Alagoas State Agency).